\documentclass[final,5p,times,twocolumn]{elsarticle}
\usepackage[ulem=normalem]{changes}
\definechangesauthor[name={Yassid Ayyad}, color={orange}]{YA}
\definechangesauthor[name={Cristina Cabo}, color={blue}]{CC}

\usepackage{amssymb}
\usepackage{amsmath}
\usepackage{url}
\usepackage{pdfpages}
\usepackage{natbib}
\usepackage{float}
\journal{Nuclear Instruments and Methods in Physics Research A}

\begin{document}

\begin{frontmatter}



\title{Two-dimensional Optical Parallel-Plate Avalanche Counter (OPPAC) with SiPM optical readout for heavy-ion tracking}


\author[IGFAE]{Yassid Ayyad}
\author[IGFAE]{Cristina Cabo}
\affiliation[IGFAE]{organization={IGFAE, Universidade de Santiago de Compostela},
                   addressline={E-15782},
                   city={Santiago de Compostela},
                   country={Spain}}

\author[FRIB]{Marco Cortesi}
\affiliation[FRIB]{organization={Facility for Rare Isotope Beams, Michigan State University},
                  city={East Lansing},
                  state={MI},
                  postcode={48824},
                  country={United States}}

\begin{abstract}

We report the progress achieved on the development of a two-dimensional Optical Parallel-Plate Avalanche Counter (OPPAC) prototype for heavy-ion tracking. The device consists of two uniform thin parallel plates of $10\times 10$~cm$^2$ effective area, separated by a gap of 3~mm. The gap is filled with Tetrafluoromethane (CF$_{4}$) at low pressure (up to 50~Torr). By applying a voltage difference between parallel plates, a uniform electric field is established within the gap. Electroluminescence emission is produced during the electron avalanche triggered by a charged particle that crosses the gas gap. The light is detected by an optical readout comprising four arrays of collimated Silicon Photomultipliers (SiPMs) deployed around the gas gap. The position of the particles is reconstructed by processing the light signals collected by all the SiPMs by computing the center of gravity of the light distribution within the SiPM arrays. The SiPM signals are read out and processed by a data acquisition (DAQ) system based on the General Electronics for TPC (GET). Preliminary results from a test with an $\alpha$-particle source and a 100 MeV/u $^{40}$Ca beam demonstrate a sub-millimetre intrinsic position resolution ($\sigma_\mathrm{det}\simeq 0.7$~mm) while preserving good linearity over the entire detector active area.
 
\end{abstract}



\end{frontmatter}

%\maketitle

\section{Introduction}
Parallel-Plate Avalanche Counters (PPACs) are widely used in nuclear reaction experiments and heavy ion accelerator facilities for beam tracking and diagnostics ~\cite{Joensen_1995, DiCarlo_2024, Wei_2020}, as they offer high time and spatial resolution at a low material budget. A PPAC detector offers a number of important features, including a homogeneous thickness of the detector, resulting in a relatively uniform loss of energy to impinging particles regardless of their position. A simple structure and relative ease of maintenance, and excellent responses to a wide array of heavy-ion varieties and energies  \cite{KUMAGAI_2013, Hanai_2023}. 

An ordinary two-dimensional position-sensitive PPAC consists of a thin central anode sandwiched between two segmented electrode foils on either side. In order to have thin pressure windows for low straggling, the detector volume is filled with low-pressure ($<$15 Torr) counting gas, generally isobutane. The normal operational condition is achieved with a voltage difference of a few hundred volts applied between the central anode and the external electrodes. The latter may consist of narrow metal strips evaporated onto 1~$\mu$m thick polymer foils, or small-diameter wires, with a pitch of around 1-2 mm. The strips on the two external electrodes are perpendicular to each other to provide two-dimensional localization capability.

Two-dimensional localization can be achieved by processing the avalanche signals captured on the strips, either by a charge-division method \cite{Cortesi_2018} or by a delay-line readout \cite{CUB_2000, DiCarlo_2024}. The charge-division method provides a wide dynamic range and reasonable stability at the expense of reduced rate capability (limited to a few kHz). Delay-line readouts can be two orders of magnitude faster (up to a few MHz), but typically have lower detection efficiency and are more prone to sporadic discharges.

A novel PPAC detector scheme based on optical readout has recently been introduced to overcome the limitations of the two readout methods described above. It is designed to provide a large dynamic range at high rates ($>$1 MHz), while maintaining a high detection efficiency~\cite{Cortesi_2018}. The detector consists of two thin parallel electrodes separated by a small gap (typically three millimeters), which is filled with low-pressure scintillating gas. Using a dedicated position-sensitive optical readout, secondary scintillation generated by avalanche processes within the gas gap can provide localization capability. The operational principle of the proposed detector was presented and discussed in detail in Ref.~\cite{Cortesi_2018}, including a systematic Monte Carlo simulation study aimed at optimizing the geometry of the photosensors for high position resolution. 

In this work, we report the first test of a two-dimensional position-sensitive OPPAC prototype under irradiation with a low-rate alpha-particle source and with a $^{40}$Ca beam. We describe the mechanical design of the detector and its main components. The present study describes the imaging performance and analysis of the image obtained using General Electronics for TPC (GET Electronics)~\cite{POLLACCO201881} as a data acquisition system (DAQ) for optical readout. Prospects and future plans are also discussed.

\section{The OPPAC prototype}  

\begin{figure*} [htpb]
  \centering
  \includegraphics [width=0.60\linewidth] {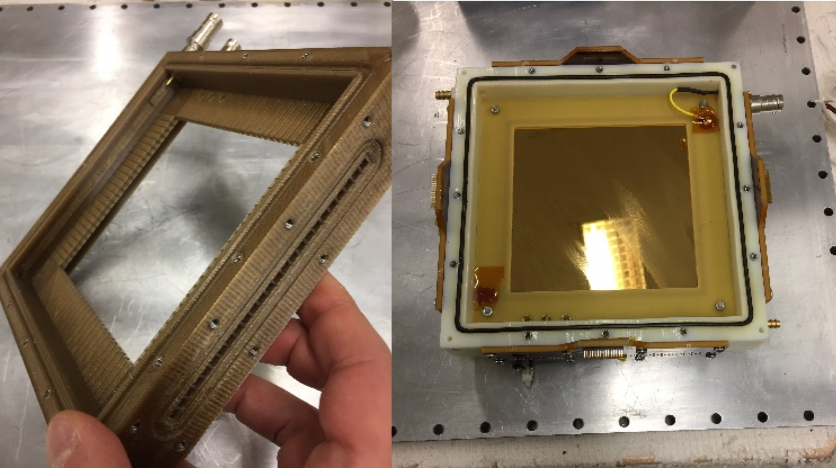}
  \includegraphics [width=0.35\linewidth] {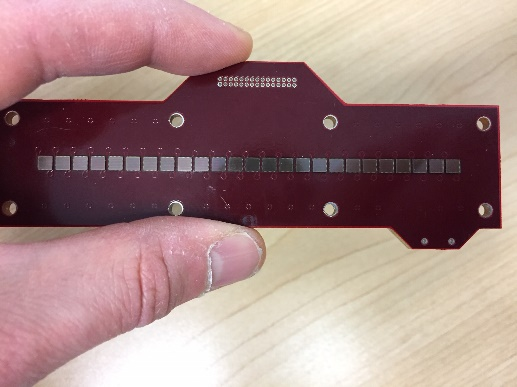}
  \caption{Upper panels: OPPAC frame fabricated in PEEK by additive manufacturing, shown empty (left) and with the polypropylene electrode foils mounted (right). Lower panel: SiPM PCB board. Four such boards, each carrying 25 SiPMs, are installed on the four sides of the OPPAC vessel.}
  \label{fig:fig_1}
\end{figure*}

The main body of the OPPAC vessel frame was manufactured with additive manufacturing processes using PEEK (polyetheretherketone). It consists of a square frame enclosing a hollow region filled with pure CF$_{4}$ at different pressures (up to 50~Torr), defining an active area of approximately $10\times 10$~cm$^{2}$ (Fig.~\ref{fig:fig_1})). Charged particles traversing the gas produce primary ionization through interactions with the gas medium. Each side of the frame contains 25 collimated square apertures of size $3\times 3$~mm$^{2}$ with a pitch of 4~mm. The length of the collimator (23.5~mm) was chosen according to the best photon collection efficiency determined by the simulation (see Ref.~\cite{Cortesi_2018}).

On the outer side of the PEEK frame, gas fittings for the gas flow and feedthroughs for high voltage and pickup signals are installed. The two PPAC electrodes were fabricated by evaporating a thin layer (150~nm) of aluminum on a stretched polypropylene foil with a thickness of approximately 0.75~$\mu$m. The electrode foils are transferred and fastened onto printed circuit boards that supply the high voltages (the cathode foil kept at high potential) and the circuitry to pick up charge signals from the PPAC (the anode foil kept at ground potential). The electrode foils enclose the PEEK frame to define a 3~mm gas gap. The detector is operated at typical electric field strength ranging from 2~kV/cm to 3~kV/cm. The strong uniform electric field established across the gas gap causes the primary electrons produced by ionization of the gas molecules to undergo avalanche multiplication, resulting in the emission of a large amount of secondary scintillation light. Aluminum frames with a $10\times 10$~cm$^{2}$ aperture, closed on both sides by aluminized Mylar foils, provide a leak-tight enclosure.

\begin{figure} [htpb]
  \centering
  \includegraphics [width=\linewidth] {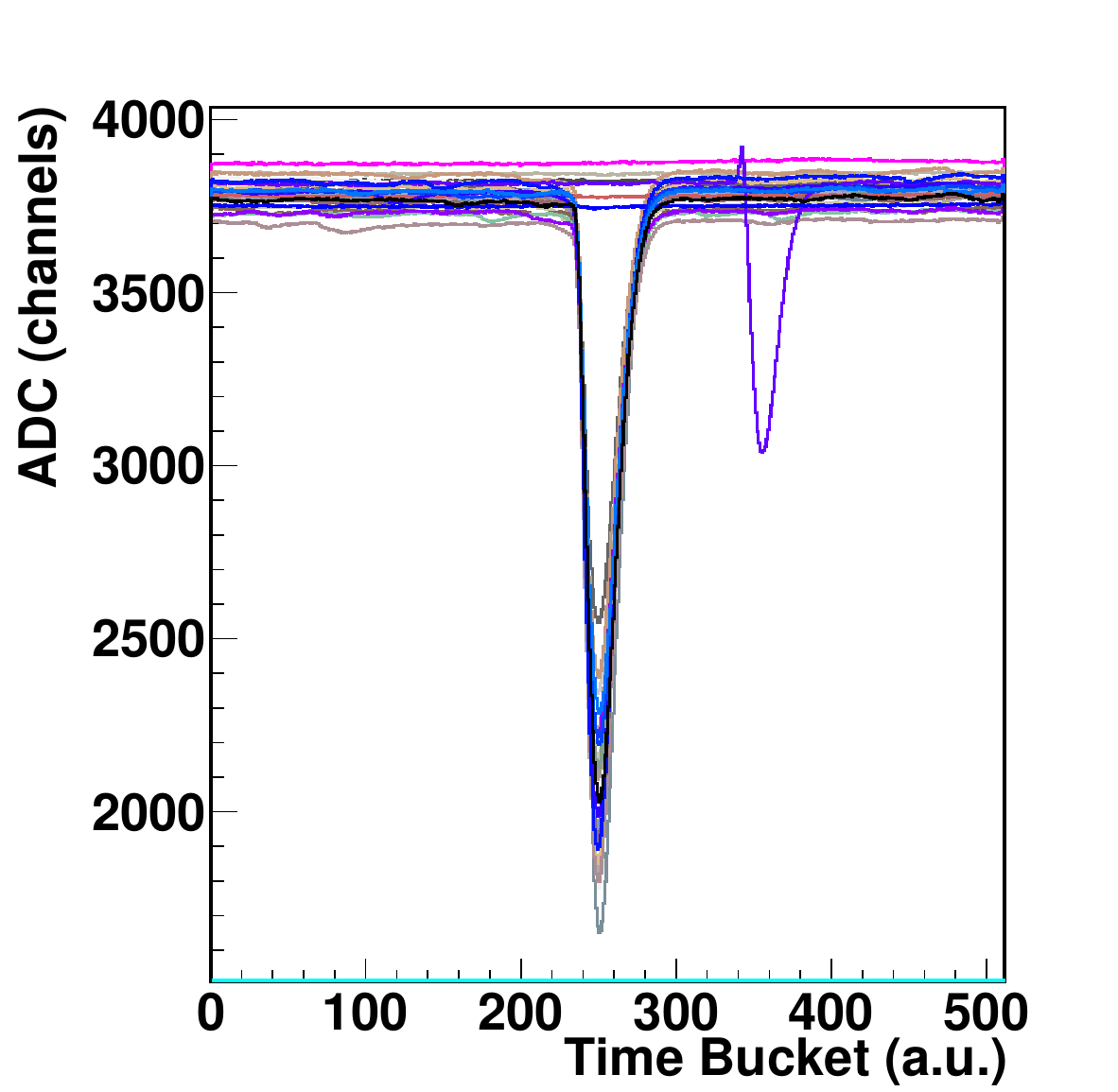}
  \caption{SiPM pulses recorded on one of the OPPAC sides, operating the detector with a field of 3300~V/cm and 30~Torr of CF$_{4}$. The electronics were configured with a gain of 10~pC, shaping time of 502~ns and a sampling frequency of 12~MHz.}
  \label{fig:fig_pulses}
\end{figure}

The optical readout for recording the secondary scintillation signals generated in the CF$_{4}$ comprises arrays of $3\times 3$~mm$^{2}$ SiPMs from Ketek~\cite{10436097}. Each side of the OPPAC frame is provided with a PCB board containing 25 SiPMs arranged in a linear configuration, with each SiPM positioned according to a collimator aperture (see Fig.~\ref{fig:fig_1}). The PCB boards are tightly secured against the frame walls with O-ring gaskets between them to ensure gas leak tightness. The interface between the SiPMs and the GET electronics is realized through an adaptor board (ZAP board) which includes protection elements to safeguard the front-end electronics. The SiPM signals are subsequently processed by the AGET chip (preamplifier$/$shaper), digitized at a configurable sampling rate ranging from 6.25 to 100 MHz, and recorded with a total of 512 time buckets. A representative example of the SiPM signals recorded on one side of the OPPAC is shown in Fig.~\ref{fig:fig_pulses}.

\section{Results}

In order to assess the performance of the OPPAC prototype in terms of localization capability, we conducted a series of systematic tests by irradiating the detector with 5.486 MeV $\alpha$-particles emitted by a $^{241}$Am source. The source was placed about 20~cm away from the detector entrance window. These tests were designed to extract the spatial resolution provided by the detector by analyzing images of a brass mask with holes of varying diameters (1~mm and 2~mm) and spacings (2.5~mm and 10~mm, respectively). The localization of the alpha-particle (along one coordinate x), based on the intensity of the electroluminescence light recorded by the SiPM arrays, is reconstructed using the procedure and algorithm explained in Ref.~\cite{Cortesi_2018}:

\begin{equation}
x = \biggl(\frac{P{_{x1}}\cdot N{_{x1}}}{\sigma_{x1}} + \frac{P{_{x2}}\cdot N{_{x2}}}{\sigma_{x2}}\biggr) \bigg/\biggl(\frac{N_{x1}}{\sigma_{x1}} + \frac{N_{x2}}{\sigma_{x2}}\biggr)
\end{equation}

where the subscripts $x1$ and $x2$ refer to the two opposite SiPM boards along the $x$ coordinate, and $P$, $N$ and $\sigma$ denote, respectively, the mean, amplitude and standard deviation of the light distribution recorded on each board. The same procedure is applied to the reconstruction of the position along the $y$ coordinate.

%\begin{figure} [htpb] 
%  \centering
%  \includegraphics [scale=0.5] {figures/mask_v2.jpg}
  %\includegraphics [scale=0.35] {figures/XY.pdf}
%  \caption{Brass mask with 1 mm holes (2.5 mm pitch), 2 mm holes (10 mm pitch) }
%  \label{fig:fig_2}
%\end{figure}

\begin{figure*} [htpb]
  \centering
  \includegraphics [width=0.45\linewidth] {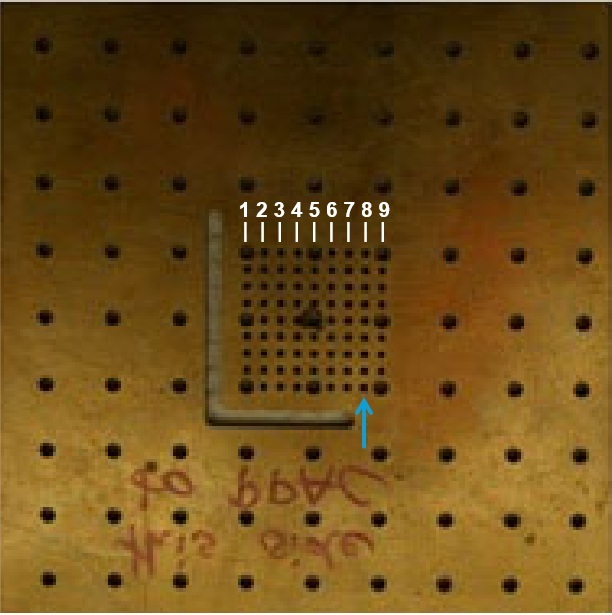}\hfill
  \includegraphics [width=0.45\linewidth] {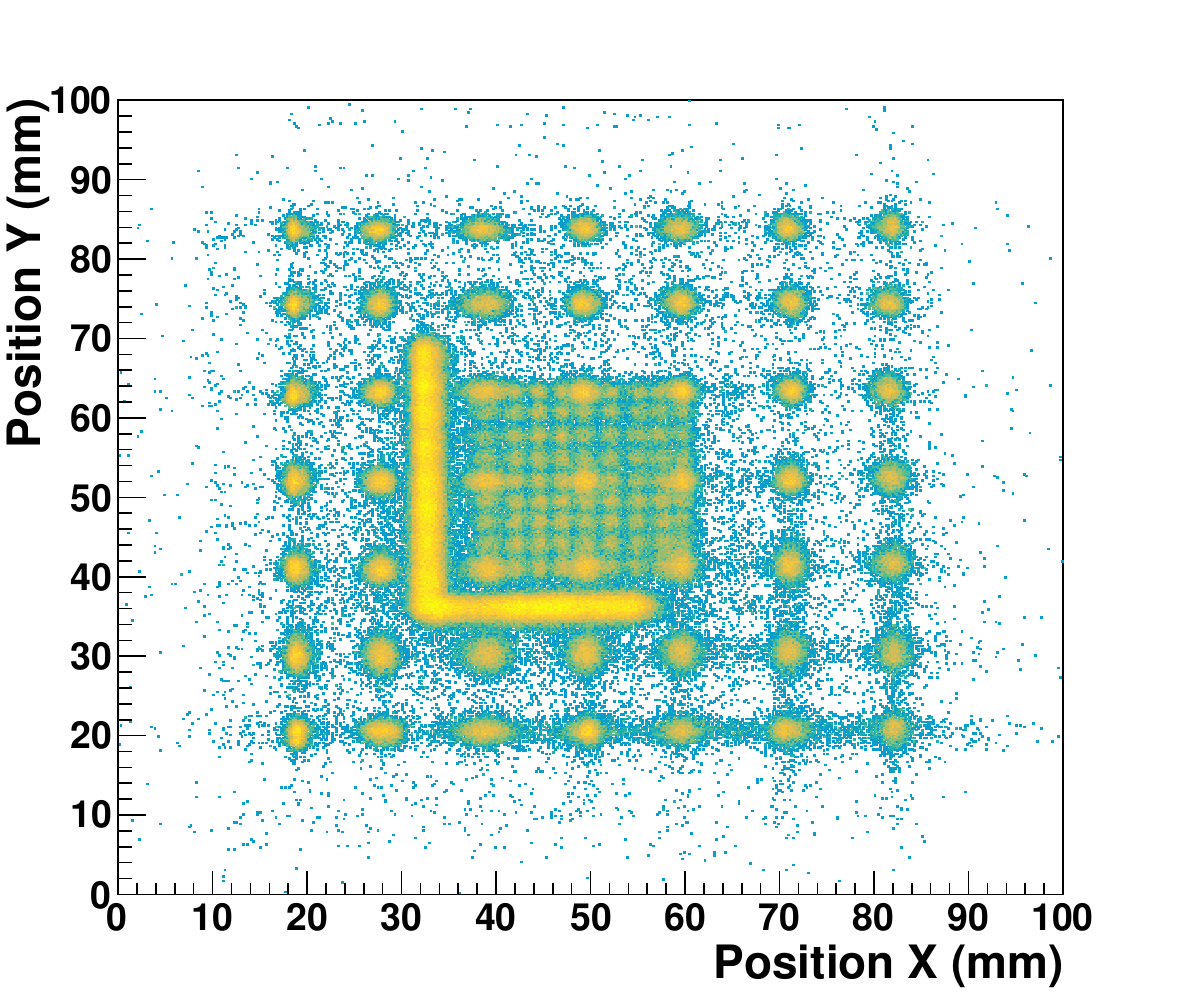}
  \caption{Brass mask used for the position-resolution measurements, featuring grids of 1~mm- and 2~mm-diameter apertures with pitches of 2.5~mm and 10~mm respectively (left), and the corresponding two-dimensional image reconstructed by the OPPAC prototype (right) under irradiation with a $^{241}$Am $\alpha$-particle source, with CF$_{4}$ at 30~Torr and an electric field of 3300~V/cm across the gas gap.}
  \label{fig:fig_3}
\end{figure*}

An example of the reconstructed image obtained with this method is illustrated in Fig.~\ref{fig:fig_3}. The figure displays the brass mask (upper panel) and the corresponding image (bottom panel) reconstructed from data recorded with the OPPAC prototype operated at a 30~Torr CF$_{4}$, with an electric field of 3300~V/cm applied across the gas gap. Each high-count spot corresponds to the reconstructed image of a single aperture, demonstrating that the OPPAC provides sufficient spatial resolution to distinguish holes separated by 2.5~mm. The reconstructed positions for the holes yield an average spacing of $2.58\pm 0.13$~mm along the X direction and $2.72\pm 0.08$~mm along the Y axis, in good agreement with the nominal 2.5~mm pitch of the brass mask. This demonstrates both the linearity of the optical readout and its ability to reproduce the physical mask geometry within a few-percent deviation.

%Figure~\ref{fig:fig_3} shows the an example of a reconstructed image counts for the nine 1 mm holes identified in Fig.~\ref{fig:fig_2}.

\begin{figure*}[htbp]
  \centering
  \includegraphics[width=0.45\linewidth]{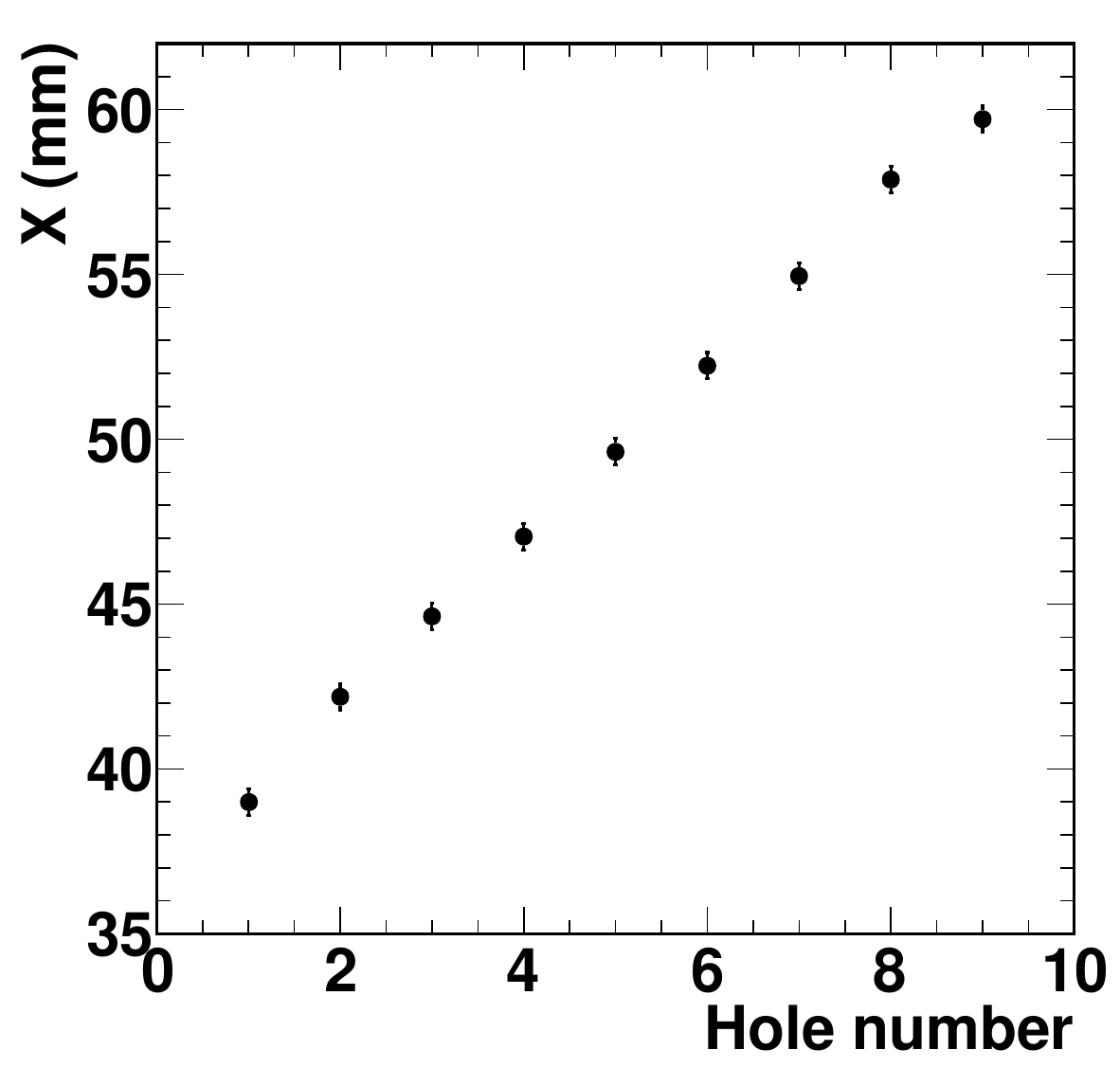}\hfill
  \includegraphics[width=0.45\linewidth]{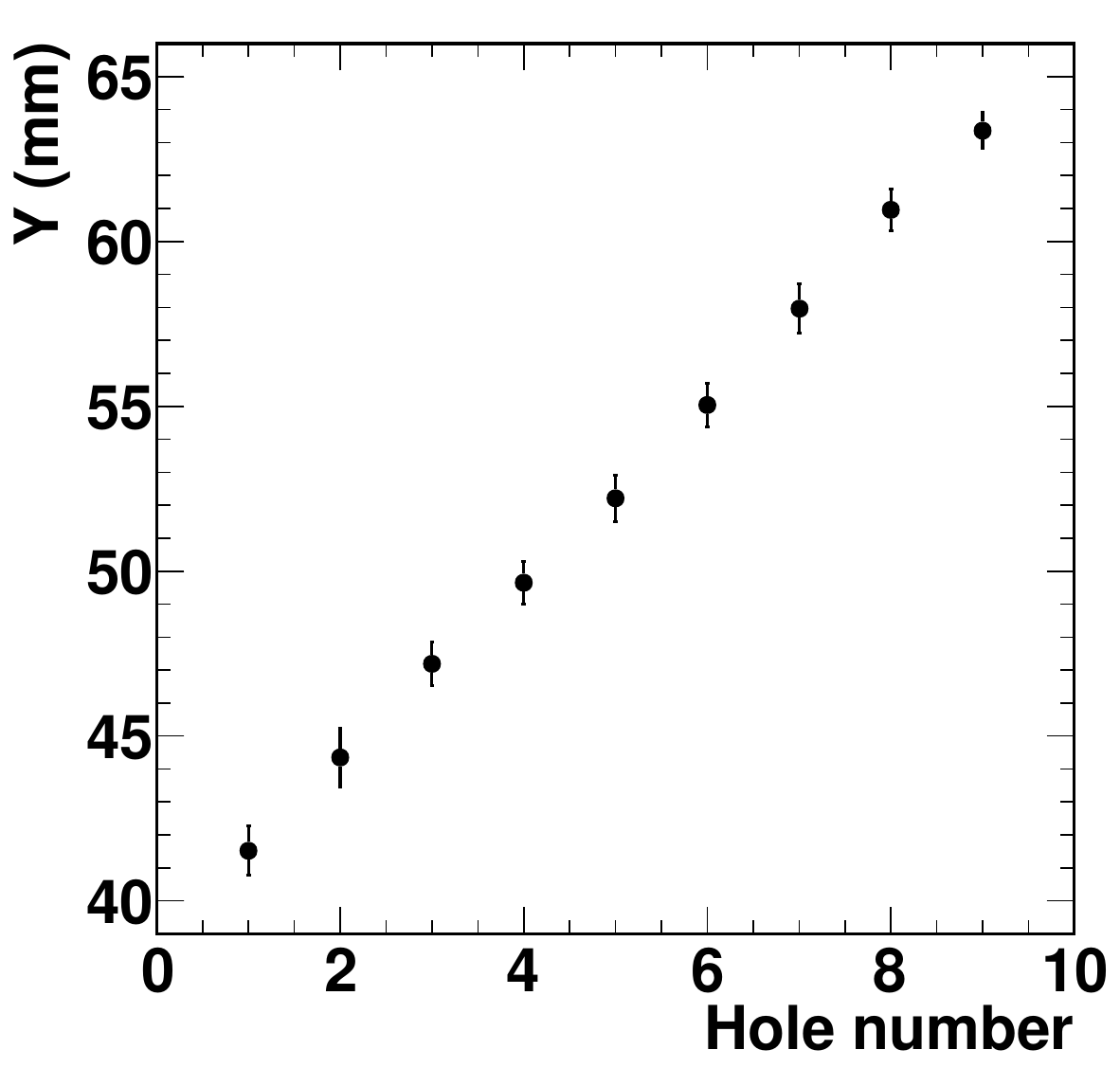}
  \caption{Reconstructed centroids along the $x$ (left) and $y$ (right) coordinates for a diagonal selection of nine 1~mm holes (numbered 1 to 9) spaced 2.5~mm in the brass mask. The Y-coordinate error bars are the Gaussian widths $\sigma$ of the individual peak fits reported in Table~\ref{tab:peak_fits}.}
  \label{fig:fig_4}
\end{figure*}

\begin{figure} [htpb]
  \centering
  \includegraphics [width=\linewidth] {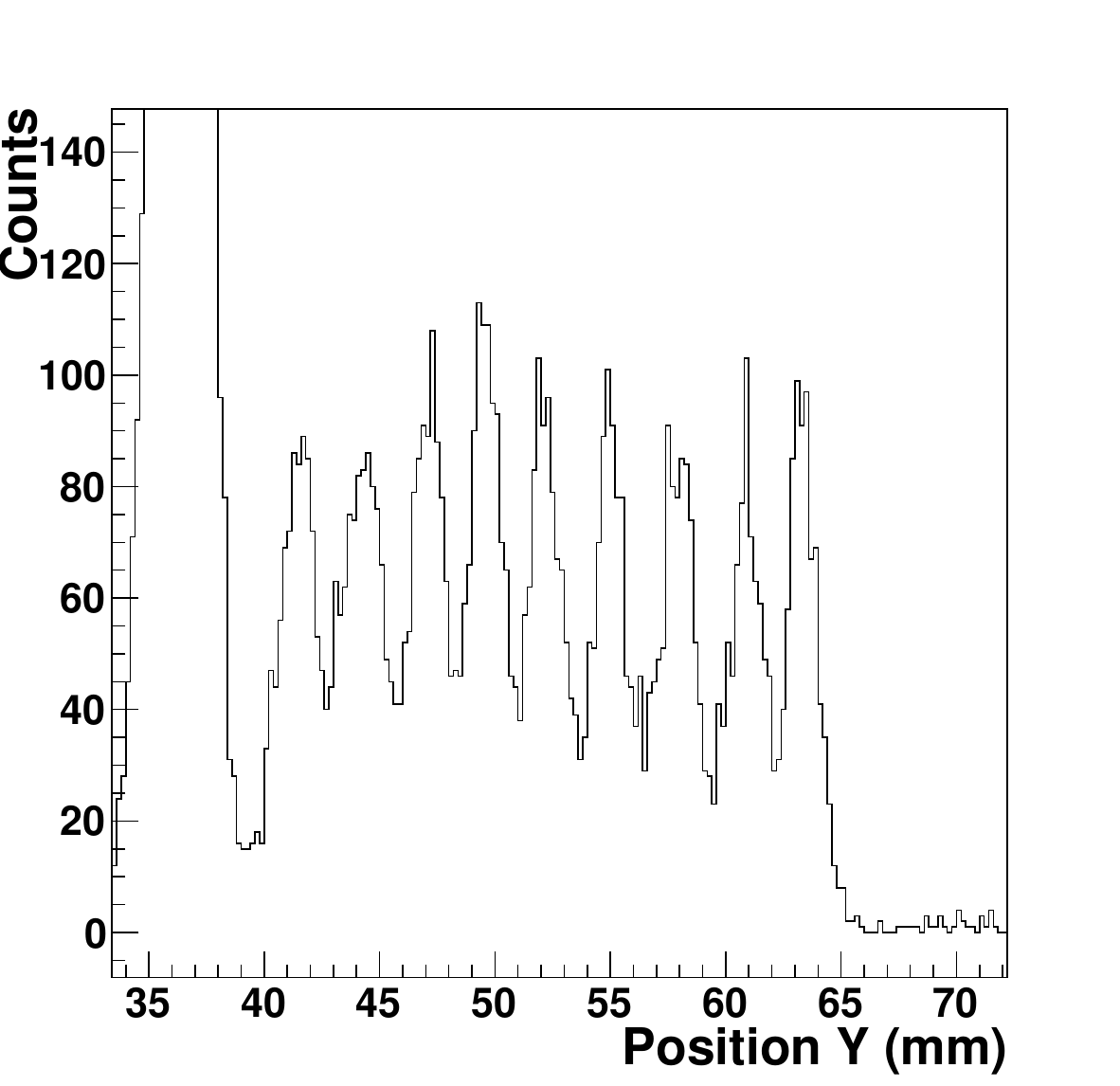}
  \caption{Projection along the $y$ coordinate of the OPPAC-reconstructed positions for the diagonal selection of nine 1~mm-diameter holes (labelled 1 to 9 in Fig.~\ref{fig:fig_4}) in the brass mask.}
  \label{fig:fig_holes}
\end{figure}

The two-dimensional reconstruction of the full brass mask, displayed in Fig.~\ref{fig:fig_3}, confirms that both the 2.5~mm and the 1~mm hole arrays are resolved simultaneously over the full $10\times 10$~cm$^{2}$ active area. The regular arrangement and uniform intensity of the reconstructed images of the mask apertures demonstrate a uniform photon collection efficiency across the field of view. No significant distortions, missing features, or edge effects were observed. The narrow, well-separated peaks of the projection in Fig.~\ref{fig:fig_holes}, together with the linear behaviour of the reconstructed centroids in Fig.~\ref{fig:fig_4}, reflect the sub-millimetre intrinsic position resolution of the prototype, in line with previous estimations from simulation studies~\cite{Cortesi_2018}.

\begin{figure}[htpb]
  \centering
  \includegraphics[width=\linewidth]{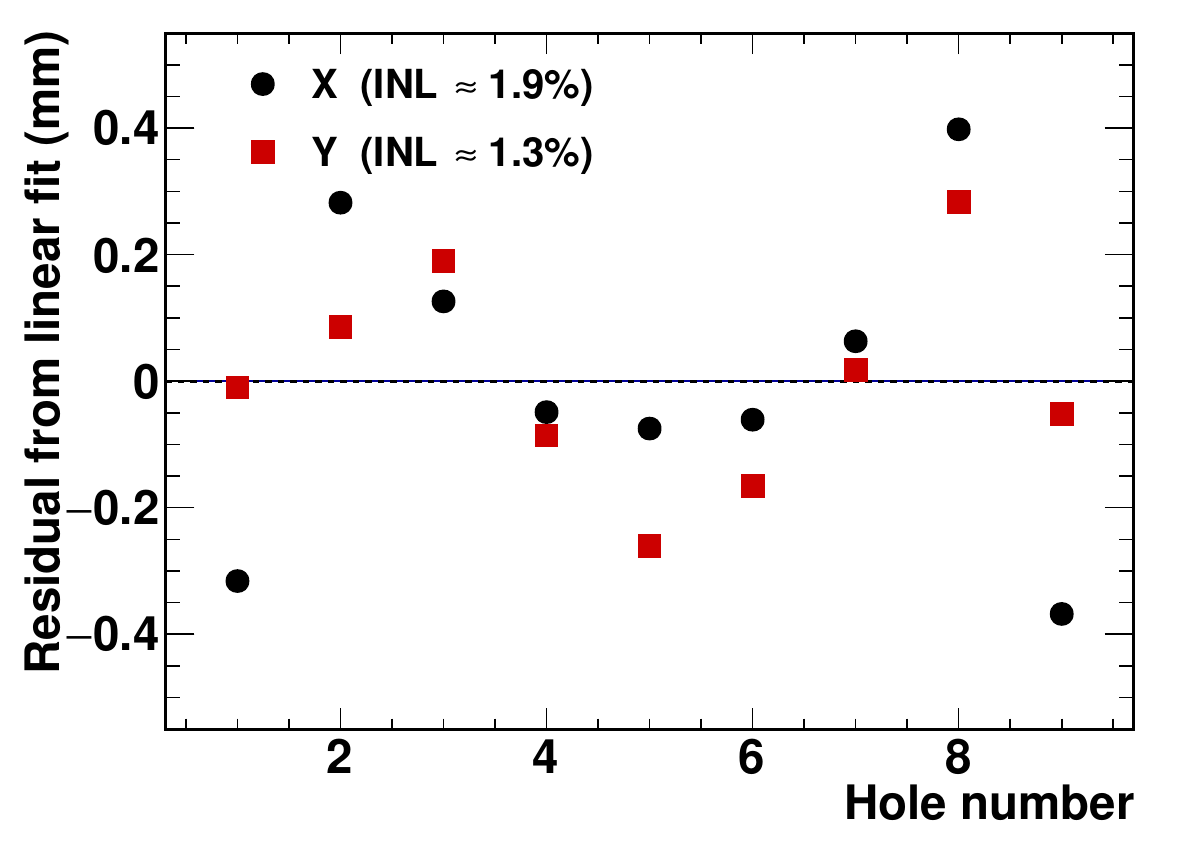}
  \caption{Residuals of the reconstructed centroids of the nine 1~mm holes (Fig.~\ref{fig:fig_4}) with respect to a linear fit of position versus hole number, for the $x$ (circles) and $y$ (squares) coordinates. The integral nonlinearity, defined as the maximum residual normalised to the full-scale range of the scan, is $\simeq 1.9\%$ in $x$ and $\simeq 1.3\%$ in $y$.}
  \label{fig:fig_inl}
\end{figure}

A quantitative measure of the imaging linearity can be extracted from the reconstructed centroids of Fig.~\ref{fig:fig_4}. For each coordinate, the nine centroids were fitted with a straight line as a function of the hole number, and the integral nonlinearity (INL) was evaluated as the maximum deviation of the measured positions from this best-fit line, normalised to the full-scale range of the diagonal scan ($\simeq 21$~mm). The resulting residuals are shown in Fig.~\ref{fig:fig_inl}: they remain within $0.40$~mm in $x$ and $0.28$~mm in $y$, with r.m.s.\ values of $0.24$~mm and $0.16$~mm, corresponding to INL figures of $\simeq 1.9\%$ and $\simeq 1.3\%$ of full scale, respectively. The residuals show no systematic trend across the scanned region, confirming that the optical centroid reconstruction remains linear at the few-percent level over the $\simeq 2$~cm diagonal probed by the 1~mm holes, with the largest excursions confined to the edge holes of the selection.

The measured peak widths are however the convolution of the intrinsic detector position resolution with the geometric profile of the $\alpha$-particle beam selected by each 1~mm mask aperture. To disentangle these two contributions, a dedicated Monte Carlo simulation was performed with Geant4~\cite{Agostinelli_2003}, reproducing the experimental geometry: the simulation modeled 5~MeV $\alpha$-particles generated from a uniform 5~mm-diameter source placed 200~mm upstream of a 1~mm-diameter aperture in a 5~mm-thick metallic collimator, with a scoring plane positioned immediately behind the aperture. Because the scoring plane lies directly downstream of the aperture, the spot at the detector is dominated by the aperture geometry rather than by the source-to-mask distance. A complementary run with the source at 400~mm yields an essentially identical distribution within Monte Carlo statistical fluctuations, demonstrating the robustness of this estimate. The spatial distribution of the $\alpha$-particles at the scoring plane is well described by the projection of a near-uniform disc of $\sim$1~mm diameter, with an r.m.s. of approximately 0.23~mm in both transverse coordinates. A Gaussian fit to the simulated one-dimensional projections, performed analogously to that applied to the experimental peaks, yields an equivalent geometric width of $\sigma_\mathrm{geom}\simeq 0.17$~mm (FWHM~$\simeq 0.4$~mm).
Assuming that the measured peak shape is the convolution of this geometric distribution with an approximately Gaussian detector response, the intrinsic position resolution can be obtained by quadratic subtraction,
\begin{equation}
\sigma_\mathrm{det}=\sqrt{\sigma_\mathrm{meas}^{2}-\sigma_\mathrm{geom}^{2}}.
\end{equation}
The nine peaks corresponding to the 1~mm holes shown in Fig.~\ref{fig:fig_holes} were individually fitted with a Gaussian peak superimposed on a linear background. The resulting per-peak parameters are summarized in Table~\ref{tab:peak_fits}. The mean fitted standard deviation across the nine peaks is $\sigma_\mathrm{meas}=0.69\pm 0.09$~mm (equivalent FWHM $\simeq 1.6$~mm), where the uncertainty corresponds to the r.m.s. spread of the individual fitted widths. Combining this value with $\sigma_\mathrm{geom}\simeq 0.17$~mm gives an intrinsic detector resolution
\begin{equation*}
\sigma_\mathrm{det}=\sqrt{\sigma_\mathrm{meas}^{2}-\sigma_\mathrm{geom}^{2}}\simeq 0.67~\mathrm{mm}\quad (\mathrm{FWHM}\simeq 1.6~\mathrm{mm}),
\end{equation*}
indicating that the geometric blur introduced by the 1~mm mask apertures contributes only at the few-percent level (in quadrature) to the measured peak widths, and that the observed spread reflects almost entirely the intrinsic position resolution of the OPPAC prototype.

\begin{table}[htbp]
  \centering
  \begin{tabular}{c c c c}
    \hline
    Peak & $\mu$ (mm) & $\sigma$ (mm) & FWHM (mm) \\
    \hline
    1 & 41.52 & 0.745 & 1.76 \\
    2 & 44.35 & 0.889 & 2.09 \\
    3 & 47.19 & 0.664 & 1.56 \\
    4 & 49.65 & 0.650 & 1.53 \\
    5 & 52.21 & 0.709 & 1.67 \\
    6 & 55.04 & 0.652 & 1.54 \\
    7 & 57.96 & 0.753 & 1.77 \\
    8 & 60.96 & 0.642 & 1.51 \\
    9 & 63.36 & 0.546 & 1.29 \\
    \hline
    Mean & --- & $0.69\pm 0.09$ & $1.64\pm 0.21$ \\
    \hline
  \end{tabular}
  \caption{Gaussian fit parameters for the nine 1~mm-hole peaks in the Y projection of Fig.~\ref{fig:fig_holes}. The uncertainties quoted on the mean are the r.m.s. spread of the individual fitted values.}
  \label{tab:peak_fits}
\end{table}

\subsection{In-beam test with a $^{40}$Ca beam at NSCL}\label{sec:beam_test}

In addition to the laboratory characterisation with the $^{241}$Am $\alpha$-particle source, the OPPAC prototype was operated under heavy-ion beam conditions during a dedicated test at the National Superconducting Cyclotron Laboratory (NSCL) at Michigan State University, where it was irradiated with a 100 MeV/u $^{40}$Ca beam. The detector was filled with pure CF$_{4}$ at a pressure of 30~Torr and biased to provide an electric field of 3300~V/cm across the gas gap, as in the $\alpha$-source measurements, and the optical readout was processed by the same GET-based DAQ. This configuration tested the operational robustness of the prototype in a high-rate, large-energy-loss environment and verified that the optical-readout scheme remains operational for heavy ions.

\begin{figure} [htpb]
  \centering
  \includegraphics [width=\linewidth] {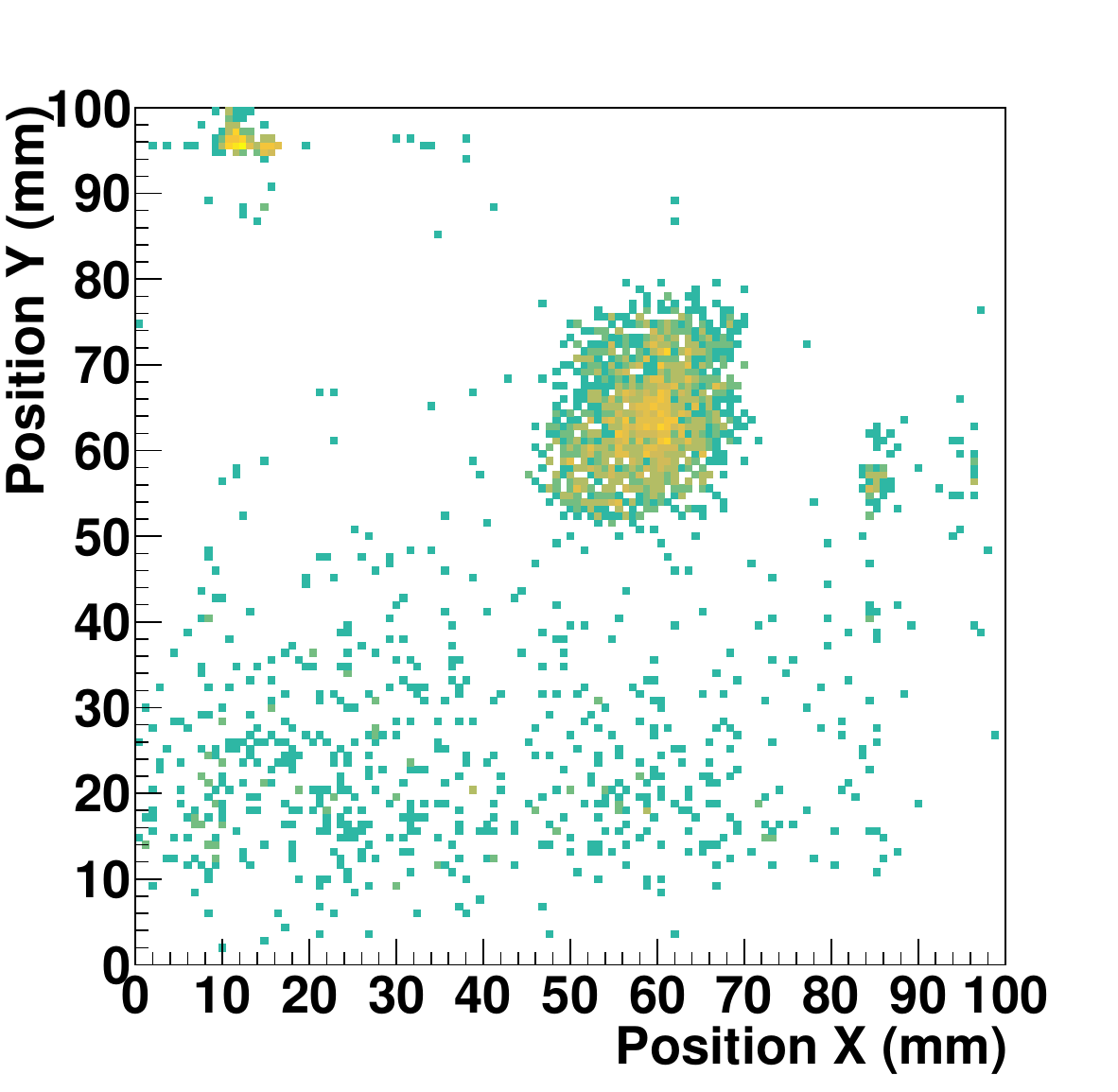}
  \caption{Two-dimensional image reconstructed by the OPPAC prototype during the in-beam test at NSCL with a $^{40}$Ca beam. The intense spot near $(x, y) \simeq (58, 64)$~mm corresponds to the primary beam, while the diffuse low-intensity events are attributed to secondaries and beam-induced background.}
  \label{fig:fig_beam}
\end{figure}

The two-dimensional image reconstructed from the SiPM signals is shown in Fig.~\ref{fig:fig_beam}. A compact, intense beam spot is clearly visible on the active area, with a centroid located at approximately $(x, y) \simeq (58, 64)$~mm and a transverse extent of roughly 20~mm in both directions, consistent with the expected beam profile delivered to the experimental area. A residual halo of low-intensity events, distributed over the lower portion of the active area and around the beam spot, is associated with secondary particles and beam-induced background. The detector operated continuously throughout the run without any sign of efficiency degradation or sporadic discharges, demonstrating that the OPPAC concept can be successfully employed for heavy-ion beam imaging and tracking under realistic experimental conditions.

\subsection{Photon yield from Garfield++/Magboltz simulation}\label{sec:phyield}

The light budget of the OPPAC is set by the secondary scintillation yield of CF$_{4}$, $Y_{\mathrm{CF}_{4}}$, defined as the average number of scintillation photons emitted per electron in the avalanche. To extract a model of $Y_{\mathrm{CF}_{4}}(E/N)$ usable for further OPPAC design studies, we have run a Garfield++/Magboltz~\cite{Garfieldpp,Biagi_Magboltz} simulation of a single electron-induced avalanche in pure CF$_{4}$, in the same uniform-field parallel-plate geometry as the OPPAC (3~mm gap, 30~Torr, 293~K). The simulation tracks every electron microscopically and reads the Magboltz per-level collision counters for all cross-section levels above a 7~eV threshold. These are converted into a \emph{direct-excitation} photon yield through a single per-excitation factor.
Because Magboltz models only direct electron-impact excitation of CF$_{4}$, this component alone does not reproduce the decreasing trend observed in the measured $Y_{\mathrm{CF}_{4}}(E/N)$. The discrepancy is generally attributed to additional scintillation channels associated with electron-ion recombination processes, which can populate excited CF$_{3}^{*}$/CF$_{4}^{*}$ states and whose contribution decreases with increasing $E/N$~\cite{Cortesi_2016}. We therefore add an empirical recombination yield of the form
\begin{equation}
Y_{\mathrm{recomb}}(E/N) = A\,\exp\!\left(-\,\frac{E/N}{E_{0}}\right),
\end{equation}
applied per avalanche electron, on top of the direct contribution.

The model has two parameters that we fix once against the parallel-plate measurement of Ref.~\cite{Cortesi_2016}: the per-excitation factor $y_{\mathrm{direct}}$ is anchored at the high-$E/N$ asymptote of that data set (where recombination is suppressed and the measured yield is dominated by direct excitation), and $(A, E_{0})$ are then chosen to fit the residual at low and intermediate $E/N$. A two-point fit on the scan endpoints yields $y_{\mathrm{direct}}\!\simeq\!2.65\!\times\!10^{-4}$~ph per Magboltz excitation, $A\!\simeq\!0.65$~ph/e$^{-}$ and $E_{0}\!\simeq\!146$~Td.

\begin{figure} [htpb]
  \centering
  \includegraphics [width=\linewidth] {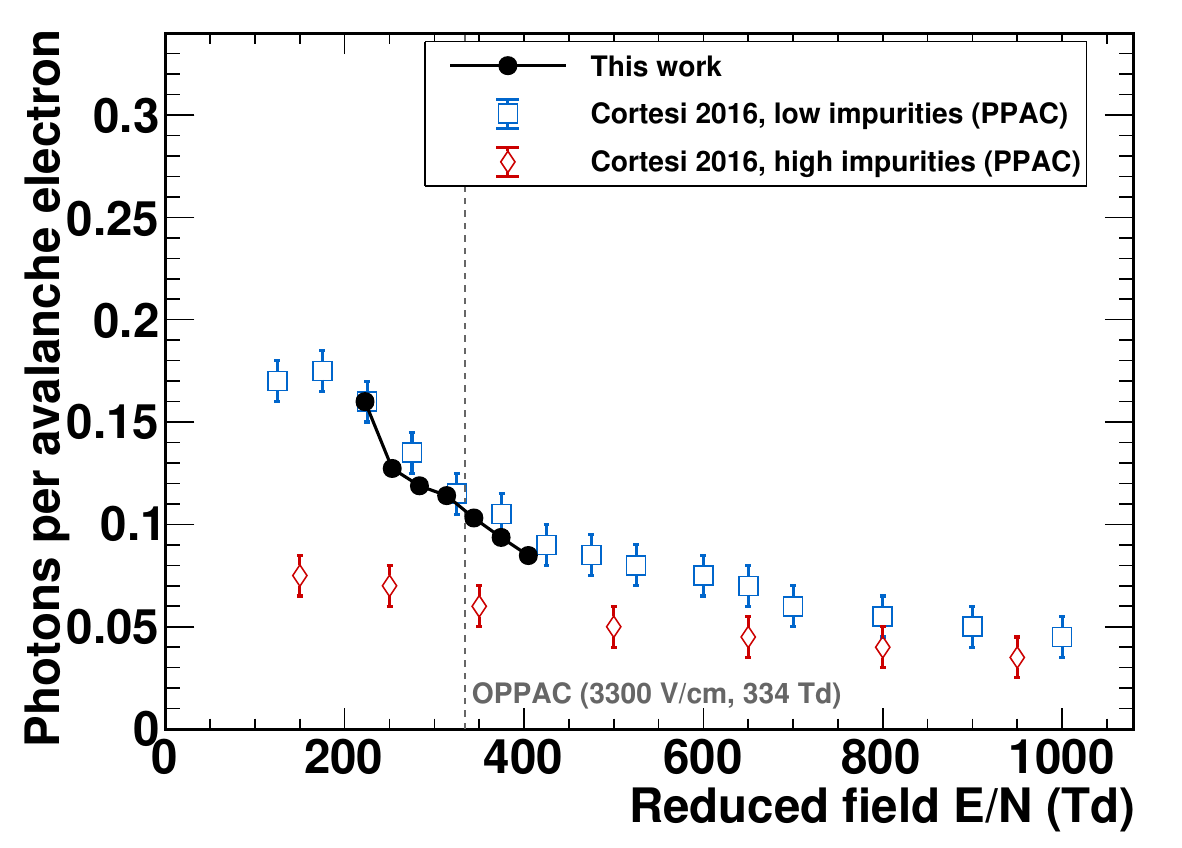}
  \caption{Secondary scintillation yield $Y_{\mathrm{CF}_{4}}$ (photons per avalanche electron) as a function of the reduced field $E/N$ in pure CF$_{4}$. Black: the present Garfield++/Magboltz simulation at 30~Torr, 293~K, in the OPPAC parallel-plate geometry, with the two-component photon model (direct Magboltz excitation plus an empirical $A\exp(-E/N/E_{0})$ recombination yield with $A=0.65$~ph/e$^{-}$, $E_{0}=146$~Td); only working points with a mean avalanche gain above 50 are shown, the regime in which the intensive ratio is well defined. Open blue squares and red diamonds: parallel-plate measurements from Ref.~\cite{Cortesi_2016} at 5--50~Torr, with and without a getter (low- and high-impurity branches, respectively); the low-impurity branch was used to fix the two free parameters of the recombination term. The vertical dashed line marks the OPPAC operating point.}
  \label{fig:fig_yield_vs_field}
\end{figure}

Figure~\ref{fig:fig_yield_vs_field} shows the resulting simulated curve overlaid on the data of Ref.~\cite{Cortesi_2016}. The two-component model reproduces the Cortesi 2016 low-impurity branch within the digitization uncertainty across the full 200--1000~Td range. At the OPPAC operating point (3300~V/cm at 30~Torr, $E/N\!\simeq\!334$~Td) the model gives $Y_{\mathrm{CF}_{4}}\!\simeq\!0.11$~ph/e$^{-}$, with $\sim$40\,\% of the yield coming from the direct-excitation channel and the remaining $\sim$60\,\% from the recombination term. This is the value adopted for the light-budget estimates in the following. Points with a simulated mean gain below 50 are excluded from the comparison. In that low-multiplication regime the avalanche is not yet fully developed: a non-negligible fraction of the counted excitations is produced by the single primary electron drifting across the gap, rather than by the multiplied electron cascade. Because the gain (the denominator of the ratio) is then small, this primary-path contribution artificially inflates $\langle n_{\gamma}\rangle/\langle n_{e}\rangle$, so the ratio no longer reflects the intrinsic excitation-to-ionization balance of a developed avalanche. We also note that the absolute avalanche gain produced by the simulation at OPPAC fields is itself capped by the absence of space-charge and photon-feedback corrections. The simulation is therefore used only for the \emph{intensive} yield ratio (photons per avalanche electron) and not for the absolute number of avalanche electrons.

A related point concerns the spectral content of the emitted light. Recent low-pressure measurements with a glass-GEM read-out~\cite{Brunbauer_2025} show that the ultraviolet emission of CF$_{4}$ (peaked near 300~nm) becomes progressively more prominent relative to the visible band (peaked near 620~nm) as the gas pressure is reduced down to a few tens of mbar, i.e.~towards the operating regime of the present prototype. In addition, an independent vacuum-ultraviolet band centred near 160~nm carries a substantial fraction of the total emission in pure CF$_{4}$~\cite{Pansky_1995}. For the present working point we take, as illustrative averages from these references, $f_{\mathrm{VUV}}\!\simeq\!0.20$, $f_{\mathrm{UV}}\!\simeq\!0.55$ and $f_{\mathrm{VIS}}\!\simeq\!0.25$ of $Y_{\mathrm{CF}_{4}}$ in the three bands.

The Ketek/Broadcom PM3315-WB SiPM mounted on the present prototype~\cite{10436097} has a peak photon-detection efficiency of $\sim$31\,\% at 430~nm at the nominal 5~V over-voltage, falling to approximately 13\,\% near 290~nm and 14\,\% near 620~nm. The silicon entrance window cuts the spectral response off below $\sim$280~nm, so the 160~nm VUV band of CF$_{4}$ is not detected. The per-band budget is summarised in Table~\ref{tab:pde_budget}. Convolving the spectral fractions with the per-band PDE values yields a spectrum-weighted effective PDE of $\langle\mathrm{PDE}\rangle\!\simeq\!0.11$. Folding this into the simulated yield gives a detected photoelectron count of $\simeq$0.012 per avalanche electron, before the geometric photon-collection factor of the four-sided SiPM array characterised in Ref.~\cite{Cortesi_2018}. The absolute number of photoelectrons per primary ionisation electron is therefore obtained by further multiplying by the experimentally measured avalanche gain and by the array solid-angle acceptance, the latter being of order a few percent for the present collimator geometry. The fact that approximately $f_{\mathrm{VUV}}\!\simeq\!20\,\%$ of $Y_{\mathrm{CF}_{4}}$ is intrinsically inaccessible to the PM3315-WB also motivates the on-going programme of SiPM-array optimisation for future OPPAC iterations, where UV-extended SiPMs (such as the VUV3-MPPC variant used by Ref.~\cite{Cortesi_2016}) or wavelength-shifting coatings could recover a sizeable fraction of the currently unseen light.

\begin{table}[htpb]
  \centering
  \caption{Photon-detection-efficiency budget for the OPPAC at the operating point ($E/N\!\simeq\!334$~Td). Columns: emission band, characteristic wavelength, fractional emission yield $f_{\mathrm{band}}$ taken from Refs.~\cite{Pansky_1995, Brunbauer_2025}, photon-detection efficiency of the Ketek/Broadcom PM3315-WB SiPM~\cite{10436097} at the band-centre wavelength and 5~V over-voltage, and the resulting contribution $f_{\mathrm{band}}\cdot\mathrm{PDE}$ to the spectrum-weighted effective PDE. The sum gives $\langle\mathrm{PDE}\rangle\!\simeq\!0.11$, used to convert the simulated $Y_{\mathrm{CF}_{4}}$ into detected photoelectrons per avalanche electron.}
  \label{tab:pde_budget}
  \begin{tabular}{lcccc}
    \hline
    Band & $\lambda$ (nm) & $f_{\mathrm{band}}$ & $\mathrm{PDE}_{\mathrm{PM3315\text{-}WB}}$ & $f\cdot\mathrm{PDE}$ \\
    \hline
    VUV & $\sim$160 & 0.20 & 0.00 & 0.000 \\
    UV  & $\sim$290 & 0.55 & 0.13 & 0.072 \\
    VIS & $\sim$620 & 0.25 & 0.14 & 0.035 \\
    \hline
    \multicolumn{4}{r}{$\langle\mathrm{PDE}\rangle =$} & 0.107 \\
    \hline
  \end{tabular}
\end{table}

\section{Future upgrades}\label{sec:future}
The intrinsic rate capability of the OPPAC concept is primarily determined by the proportional-avalanche dynamics in the PPAC gas gap and by the recovery time of the SiPMs. The achievable global counting rate depends strongly on the transverse beam size, since the dominant limitation arises from the probability of avalanche overlap within the active area. For beam spots extending over areas larger than $\sim 50$~cm$^{2}$, operation at rates exceeding 1~MHz is expected to be sustainable. Modern SiPMs exhibit microcell recovery times of the order of a few tens of nanoseconds, allowing sustained local hit rates well above 1~MHz with only moderate saturation effects.

In the present prototype, however, the achievable instantaneous rate is limited by the read-out chain: the GET electronics used in this work is optimised for time-projection-chamber-like applications, where each event is recorded as a long time-sampled waveform (512 buckets per channel), and its trigger and dead-time structure constrain the effective throughput to a few kHz of complete two-dimensional events. This is fully adequate for the laboratory characterisation reported here and for low-rate beam diagnostics, but it would become the dominant bottleneck for applications such as in-flight tracking at heavy-ion facilities, where local instantaneous rates of $10^{5}$--$10^{6}$~Hz per detector are routinely required.

A natural path towards a scalable, large-area version of the detector is to migrate the optical read-out to the Scalable Readout System (SRS) developed by the former RD51 collaboration~\cite{Martoiu_2013} (currently known as DRD1). The SRS is a modular, FPGA-based architecture conceived for micro-pattern gaseous detectors, with several front-end ASIC options interfaced through Front-End Concentrator (FEC) cards to a configurable trigger and data-aggregation layer. It has become the \emph{de facto} standard read-out for MPGD-based trackers (GEM, Micromegas, $\mu$RWELL...), so an SRS-based OPPAC integrates naturally with the existing tooling, trigger logic and analysis frameworks already in use at large heavy-ion facilities such as FRIB and CERN.

Among the available front-end ASICs, the VMM3a is particularly well suited to the OPPAC optical read-out. Each VMM3a chip provides 64 self-triggered channels with adjustable shaping time in the $\sim$25--200~ns range (well matched to the fast SiPM and CF$_{4}$ scintillation signals) together with per-channel charge digitisation and time-stamping at the nanosecond level. The 100 SiPMs of the present prototype can be instrumented by a single 128-channel VMM3a hybrid, with comfortable margin for future channel-count expansion. A dedicated high-dynamic-range SiPM front-end adapter for the VMM, providing input matching, SiPM bias and amplitude scaling tailored to the direct coupling of SiPMs, has recently been developed within the RD51 collaboration~\cite{Rusu_RD51_2023} and further simplifies this integration.

Compared to the GET system, an SRS/VMM3a read-out replaces the full 512-bucket sampled waveform with the per-channel charge and timestamp. This is fully sufficient for the centre-of-gravity reconstruction of Eq.~(1), which uses amplitudes only, and reduces the per-event data volume by more than an order of magnitude. The total light yield summed over the four SiPM sides, which scales with the energy deposited in the gas gap, also remains accessible.

From the system point of view, the SRS architecture scales linearly with the number of FEC cards, so larger OPPAC active areas, denser SiPM pitches, or multi-OPPAC focal-plane configurations can be instrumented without re-designing the DAQ. Taken together, these features make an SRS-based read-out a natural route towards a scalable, large-area implementation of the OPPAC concept, and the corresponding developments are currently underway.

\section{Summary and Conclusions}

A first test of a two-dimensional Optical Parallel-Plate Avalanche Counter (OPPAC) prototype has been successfully carried out using $\alpha$ particles from a $^{241}\mathrm{Am}$ source, complemented by an in-beam test with a $^{40}$Ca beam at NSCL. The detector, operated with CF$_{4}$ gas at low pressure, employed a four-sided SiPM optical readout coupled to a GET-based acquisition and processing system. This configuration allowed the optical collection and digitization of the electroluminescence generated during the avalanche process, enabling two-dimensional reconstruction of the particle interaction points within the active area.

The detector response was evaluated using a brass mask featuring two grids of circular apertures with diameters of 1~mm and 2~mm, and pitches of 2.5~mm and 10~mm, respectively. The reconstructed images obtained from the SiPM arrays demonstrated that the OPPAC prototype can accurately reproduce the mask geometry. The measured mean pitch values of $2.58\pm 0.13$~mm in the X direction and $2.72\pm 0.08$~mm in the Y direction are consistent with the nominal 2.5~mm spacing, confirming both the linearity of the optical readout and the correct spatial calibration of the reconstruction algorithm. Furthermore, individual Gaussian fits to the reconstructed 1~mm-hole peaks yield a mean standard deviation of $0.69\pm 0.09$~mm (FWHM $\simeq 1.6$~mm). After deconvolving the geometric contribution of the 1~mm mask apertures, estimated with a dedicated Geant4 simulation, the intrinsic position resolution is $\sigma_\mathrm{det}\simeq 0.67$~mm, demonstrating a homogeneous, sub-millimetre response across the entire $10\times 10$~cm$^{2}$ active area.

The in-beam test carried out at NSCL with a $^{40}$Ca beam further confirmed the operational robustness of the prototype under realistic experimental conditions: the beam spot was clearly imaged on the active area (Fig.~\ref{fig:fig_beam}) and the detector operated continuously without any sign of efficiency degradation or sporadic discharges, demonstrating the applicability of the optical-readout scheme to heavy-ion beam tracking.

The results obtained validate the feasibility and robustness of the optical readout concept for PPAC-type detectors. The OPPAC prototype demonstrates that position information can be reliably extracted from electroluminescence light, eliminating the need for conventional charge readout structures such as strips or delay lines. This approach preserves the low material budget characteristic of PPACs, while offering additional advantages such as a simplified mechanical design, flexibility in sensor layout, and potential operation at higher particle rates.

Future work will focus on a more detailed characterisation of the detector response under different CF$_{4}$ pressures and electric field configurations, with the goal of optimizing light yield and spatial resolution. As discussed, the rate capability of the present prototype is set by the GET-based read-out rather than by the detector itself. A migration of the SiPM read-out to the Scalable Readout System (SRS) is foreseen to provide a modular, scalable read-out for large-area, many-channel OPPAC systems, while dedicated fast front-end electronics would be required to fully exploit the intrinsic timing and high-rate capability of the detector. Further dedicated tests with heavy-ion beams are planned to evaluate the high-rate behaviour and long-term operational stability of the detector under realistic experimental conditions. These studies will establish the scalability of the OPPAC concept for large-area tracking detectors and its suitability for beam tracking and vertex reconstruction in nuclear physics experiments.

\section*{Acknowledgments}
This material is based upon work that was partially supported by the U.S. Department of Energy, Office of Science, Office of Nuclear Physics and user resources of the Facility for Rare Isotope Beams (FRIB) Operations, which is a DOE Office of Science User Facility under Award Number DE-SC0023633.

This work has received financial support from "Cátedra Televés en Diseño Microelectrónico" (TSI-069100-2023-0010) by the PERTE Chip, Secretaría de Estado de Telecomunicaciones e Infraestructuras Digitales, Ministerio para la Transformación Digital y de la Función Pública and has been co-funded by the European Union-NextGenerationEU; and from the Xunta de Galicia (CIGUS Network of Research Centres) and the European Union. We thank the support from the “María de Maeztu” Grant CEX2023-001318-M, funded by MICIU/AEI/ 10.13039/501100011033.

Y.A. is supported by the Spanish Ministerio de Economía y Competitividad through the Programmes “Ramón y Cajal” with the grant number RYC2019-028438-I funded by MCIN/AEI/10.13039/501100011033. Y.A. is supported by the Spanish Ministry of Science and Innovation through the Proyectos de Generación de Conocimiento 2021 grant PID2021-125995NA-I00 and by the Regional Government
of Galicia under the program “Proyectos de excelencia” Grant No. ED431F 2022/13.

\bibliographystyle{unsrt}
\bibliography{bibliography}

\end{document}